\documentclass[usenatbib]{mnras}

\usepackage{newtxtext,newtxmath}
\usepackage{braket}
\usepackage{caption}
\usepackage{subcaption}
\usepackage{booktabs}
\usepackage{multirow}
\usepackage{tabularx}
\usepackage{amsmath}

\usepackage[T1]{fontenc}

\DeclareRobustCommand{\VAN}[3]{#2}
\let\VANthebibliography\thebibliography
\def\thebibliography{\DeclareRobustCommand{\VAN}[3]{##3}\VANthebibliography}

\usepackage{graphicx}	
\usepackage{amsmath}	
\usepackage[dvipsnames]{xcolor}
\usepackage{ulem}
\usepackage{cancel}

\newcommand{\msun}{~\rm M_{\large \odot}}
\newcommand{\lsim}{\mathrel{\hbox{\rlap{\lower.55ex\hbox{$\sim$}} \kern-.3em\raise.4ex\hbox{$<$}}}}

\title[Optical follow-up of tick-tock MBHB candidate]{Optical follow-up of the tick-tock massive black hole binary candidate}

\author[Dotti et al.]{Massimo Dotti$^{1,2,3}$\thanks{massimo.dotti@unimib.it}, Matteo Bonetti$^{1,2,3}$\thanks{matteo.bonetti@unimib.it}, Fabio Rigamonti$^{2,3,4}$, Elisa Bortolas$^{1,2}$, Matteo Fossati$^{1,3}$,
\newauthor Roberto Decarli$^5$, Stefano Covino$^{3}$, Alessandro Lupi $^{1,2,5}$,
Alessia Franchini$^{1,2}$, Alberto Sesana$^{1,2,3}$, 
\newauthor Giorgio Calderone$^6$
\\
$^{1}$Dipartimento di Fisica ``G. Occhialini'', Universit\`a degli Studi di Milano-Bicocca, Piazza della Scienza 3, I-20126 Milano, Italy\\
$^{2}$INFN, Sezione di Milano-Bicocca, Piazza della Scienza 3, I-20126 Milano, Italy\\
$^{3}$INAF - Osservatorio Astronomico di Brera, via Brera 20, 20121 Milano, Italy\\
$^{4}$DiSAT, Universit\`a degli Studi dell'Insubria, via Valleggio 11, I-22100 Como, Italy\\
$^{5}$INAF – Osservatorio di Astrofisica e Scienza dello Spazio di Bologna, Via Gobetti 93/3, I-40129 Bologna, Italy\\
$^6$INAF – Osservatorio di Astronomico di Trieste
Via G.B. Tiepolo, 11, I-34143 Trieste, Italy\\
}

\date{Accepted 2022 November 12. Received 2022 November 12; in original form 2022 May 12}

\pubyear{2022}

\begin{document}
\label{firstpage}
\pagerange{\pageref{firstpage}--\pageref{lastpage}}
\maketitle

\begin{abstract}
The observation of a population of massive black hole binaries (MBHBs) is key for our complete understanding of galaxy mergers and for the characterization of the expected gravitational waves (GWs) signal.
However, MBHBs still remain elusive with only a few candidates proposed to date. Among these, SDSSJ143016.05+230344.4 ('tick-tock' hereafter) is the only candidate with a remarkably well sampled light curve showing a clear reduction of the modulation period and amplitude over three years of observations. This particular feature has been recently claimed to be the signature of a MBHB that is about to merge.
In this paper, we provide an optical follow-up of the tick-tock source using the Rapid Eye Mount (REM) telescope. 
The decreasing luminosity observed in our follow up is hardly explained within the binary scenario. We speculate about an alternative scenario that might explain the observed light curve through relativistic Lense-Thirring precession of an accretion disc around a single massive black hole.
\end{abstract}



\section{Introduction}

The formation of massive black hole (MBH, i.e. black holes with masses in the range $10^5-10^{10}\,M_{\odot}$) pairs in a single galaxy are the natural outcome of the galaxy mergers \citep[e.g.][]{BBR80,vw12, 2013CQGra..30x4008M,steinborn,volonteri16,rg19,Derosa2019}, and are expected to be the progenitors of extremely loud sources of low-frequency gravitational waves \citep[see][and references therein]{PTA, AmaroSeoane2017,2022arXiv220306016A}. Some of these pairs can shrink their separation due to dynamical friction exerted by the diffuse gas and star distribution of the merger remnant down to the formation of a bound MBH binary \citep[MBHB; e.g.][]{review}, where they evolve towards the gravitational wave (GW) emission stage due to interactions with  stars \citep[e.g.][]{2018MNRAS.474.1054B,2018MNRAS.477.2310B} and gas \citep[e.g.][]{2021ApJ...918L..15B,2021MNRAS.507.1458F,2022ApJ...929L..13F}. The fraction of pairs that evolve in MBHBs is still unclear, and it depends critically on the MBH pair mass ratio and on the dynamical and morphological properties of the galaxy merger remnant  \citep{callegari,fiacconi,delvalle,2020ApJ...899..126S,bortolas20,bortolas22}.

The direct observation of a population of MBHBs could significantly improve our understanding of MBH pairing in galaxy mergers and better constrain the expected GW signals in current and future gravitational wave searches.
To date hundreds of MBHB candidates have been published, either as resolved pairs of flat-spectrum radio cores \citep[0402+379, with a projected separation of 7 pc; see][]{rodriguez09},  or through the search of asymmetric/shifted broad emission lines \citep[][]{tsalmantza,eracleous} and periodic varying light-curves  \citep{valtonen, Ackermann15, Graham15, Li2016, Charisi16, Sandrinelli16,Sandrinelli18, Severgnini18, Li+2019, LiuGez+2019,Chen+2020}.

In particular, the latter has been associated to MBHBs with separations $\lsim 0.01$ pc, as they are too compact to retain individual broad line regions \citep[depending on the observed transition, the MBH masses and their luminosities, see][]{montuori11,montuori12}.

In general, periodic varying light-curves have been interpreted either as the imprint of periodic feeding events onto the MBHs due to the torque exerted by the binary onto the surrounding gas \citep[see the recent studies in][and references therein]{bowen,Dascoli}, or as the effect of a secondary plunging into the accretion disc of the primary on very eccentric orbits \citep[][]{valtonen}, or as the Doppler boosting associated with the motion of one of the binary component at similar scales \citep[e.g.][]{dorazio}.

The number of MBHBs detectable through light-curve modulation is expected to be a decreasing function of the MBHB separation, due to the decreasing residence time \citep[e.g.,][]{xin}.  
SDSSJ143016.05+230344.4 \citep[dubbed `tick-tock' hereafter; see][]{ticktock} is the only candidate proposed to date showing a drastic evolution of the modulation period, that decreased from $\sim 1$ yr down $\sim 1$ month over three years of observations. The MBHB post-Newtonian (PN) model proposed in the discovery paper requires the pericentre of the binary to be $\sim 10$ gravitational radii, and predicts the binary to coalesce within three years.

It must be stressed that alternative explanations have been proposed for all the classes of MBHB candidates (with the possible exception of 0402+379). Individual spectroscopy--based candidates, for example, have been interpreted as chance super-positions of two galaxies within the angular resolution of the spectrograph used \citep[e.g.][]{heckman}, as recoiling MBHs due to anisotropic GW emission at coalescence \citep[e.g.][]{komossa}, or as disc-like broad line regions around a single MBH  \citep[][]{DPE,tsalmantza}. Indeed, some of the proposed objects have already been proven not to be MBHBs \citep[][]{decarli09a,decarli09b,decarli14}, and only few systems keep showing an evolution of the line profiles consistent with MBHBs with orbital periods of few decades \citep[][]{runnoe17}.  Alternative interpretations have been proposed for quasi-periodically modulated sources as well as, either assuming Lense-Thirring driven precessing jets \citep[][]{Sandrinelli16,lt} or only red-noise in the variability of AGN \citep[][]{vaughan, liu}.

Since compact MBHBs are also loud low-frequency gravitational wave (GW) emitters \citep[e.g.][]{2013CQGra..30x4009S}, periodic candidates can also be constrained using pulsar timing array (PTA) data \cite[see, e.g.,][for a proof of concept]{2004ApJ...606..799J}. Although, as of today, all candidates are individually consistent with PTA upper limits, when collectively considered, their parent population has been shown to be in moderate \citep[when considering candidate selected in the optical bands,][]{2018ApJ...856...42S} to severe \citep[when considered X-ray selected blazar candidates,][]{2018MNRAS.481L..74H} tension with those limits. It should be noticed, however, that the PTA collaborations recently reported the presence of a common red signal slightly larger than some of the previously published upper limits \citep{2020ApJ...905L..34A,2021ApJ...917L..19G,2021MNRAS.508.4970C,2022MNRAS.510.4873A}. Should the GW origin of this signal be confirmed, it would alleviate the tension at least with optically selected candidates.

In this study we aim at testing the binary hypothesis for tick-tock. We start by presenting two new optical follow-ups for the candidate (section~\ref{sec:observations}). We then describe the MBHB model developed in \citet{ticktock}, including our corrections to it (section~\ref{sec:binary}), and  we comment on how these new observations combined with the already available observational data constrain the tick-tock MBHB model (section~\ref{sec:tests}). The caveat of the current analysis are discussed in section~(\ref{sec:caveats}). 
Finally, in section~(\ref{sec:conclusions}) we present our conclusions. 

\section{Optical follow-up}\label{sec:observations}

\begin{figure*}
    \centering
    \includegraphics[scale=0.6]{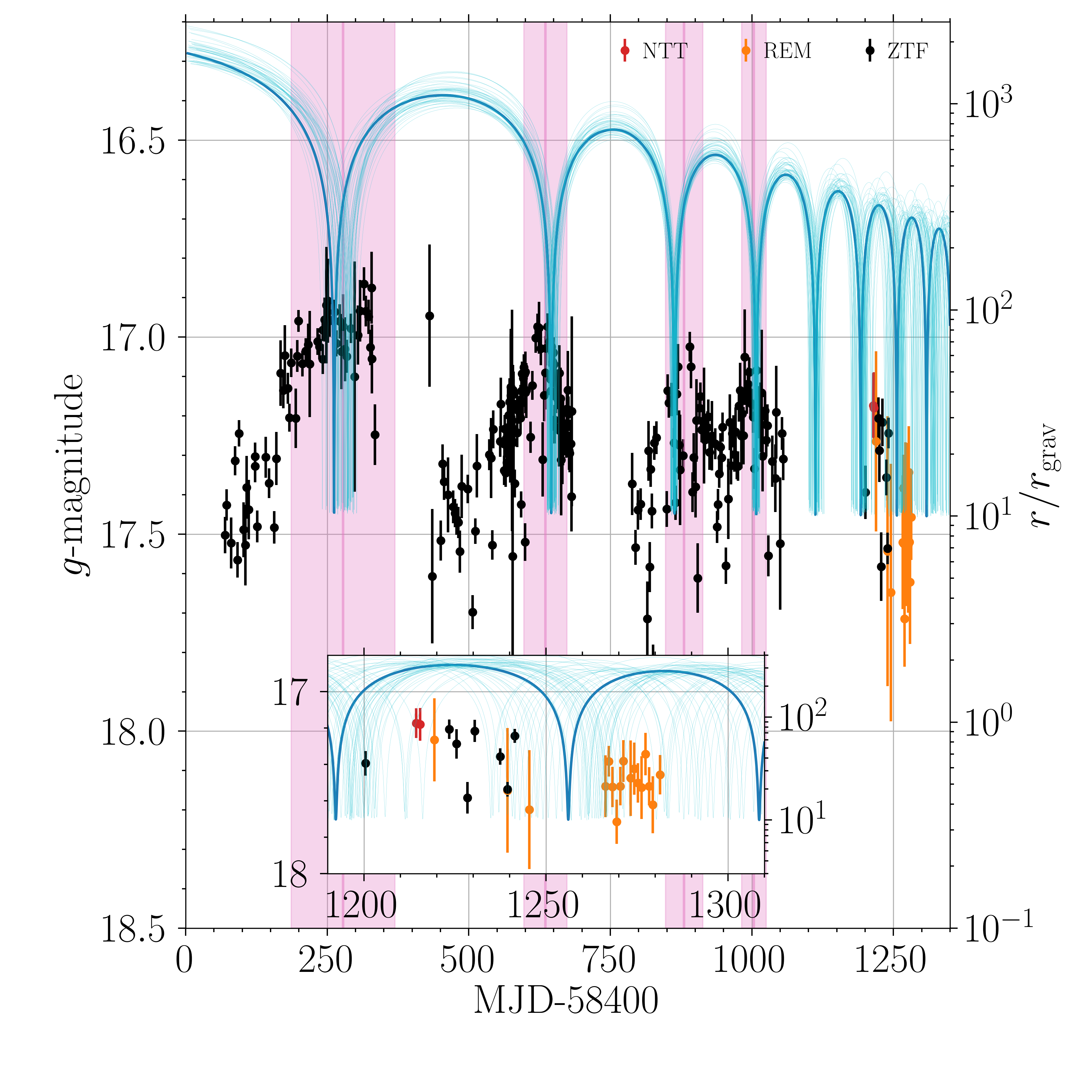}
    \caption{Light curve of the tick-tock source in the $g$-band (points with error bars), with superimposed the PN evolution of the binary separation best matching the four ZTF optical peaks. The blue line is the maximum posterior while the cyan lines represent the 50 most likely realizations. The pink vertical lines (shaded areas) mark the peak estimates (1-$\sigma$ Gaussian uncertainties) provided by \citet{ticktock}. Black, red and orange points refer to the ZTF, NTT and REM $g$-magnitudes respectively, all extracted within 3$''$ (see text).
    The inset shows a zoom in on the new optical follow-ups. The orbital separation axis is in logarithmic scales to best highlight the close to constant values of the pericentres, a general feature of GW driven orbital decay.}
    \label{fig:photometry}
\end{figure*}

We set up a monitoring campaign of tick-tock to measure its optical broad-band photometry over $\sim 10$ weeks following the publication of the \citet{ticktock} paper.

The first two observations of tick-tock have been performed using the ESO Faint Object Spectrograph and
Camera \citep[EFOSC2;][]{efosc2,efosc2b} on the
ESO/New Technology Telescope (NTT) on February 4 and 5, 2022.
On the first night, the weather was rather unstable, and the
seeing in both g and r during the observations was ~1.7". On the second
night, we observed with stable, photometric conditions. The seeing was
0.8". The airmass was 2.0 during the February 4 observations, and 1.7 on
February 5. In both visits, we observed the source in $6\times 1$ min
frames in g and r (Gunn filters g\#782 and r\#784 on EFOSC2), with small
$<15''$ shifts for dithering. The data were processed with our custom
suite of IRAF\footnote{IRAF is distributed by the National Optical
Astronomy Observatory, which is operated by the Association of
Universities for Research in Astronomy (AURA) under a cooperative
agreement with the National Science Foundation.} routines. The
astrometry package by L.~Wenzl\footnote{https://github.com/lukaswenzl/astrometry/} secured our
astrometric calibration. 

The remaining photometric points have been obtained with the Rapid Eye Mount telescope (REM), a Ritchey-Chretien configuration with a 60 cm primary mirror and a total f/8 focal ratio. It is a fully robotic telescope located at the ESO La Silla Observatory \citep[][]{zerbi,covino}  which hosts the REMIR near-infrared camera \citep[][]{conconi} and the ROSS2 optical camera \citep[][]{tosti}. The two instruments observe the same field of view of 10 × 10 arcmin almost simultaneously. The REMIR camera has a rotating wheel that accommodates three filters, J, H, K. The ROSS2 camera allows for simultaneous imaging in the SDSS $g$, $r$, $i$, $z$ filters \citep{York00} thanks to a band splitting obtained with plate dichroics. We will limit our analysis to the g-band magnitudes, where the largest fluctuations are observed  \citep[the redder bands ($r$,$i$,$z$ and NIR) are increasingly strongly contaminated by the host light which is dominated by red stellar populations, as observable also in the spectra reported in][]{ticktock}.
Data reduction was carried out following the standard procedures, with the subtraction of an averaged bias frame dividing by the normalized flat frame. 

Both for NTT and REM observations the zero point of the images have been computed by comparing the instrumental photometry of field stars with the SDSS catalog magnitudes. 

Differently from the Zwicky Transient Facility (ZTF) data analysis presented in \citet[][]{ticktock}, we did not measure PSF-photometry. We kept instead a fixed 3$''$ radius aperture. In order to have a meaningful comparison between the different data sets we re-evaluated the g magnitudes in few ZTF images within 3$''$ adopting the following procedure. First we downloaded ZTF postage-stamp images that contribute to the light-curve provided by the ZTF DR11. From each image, we then extracted the flux of the source in a 3$''$ radius aperture and the flux of a calibration star (SDSS $g$ = 16.51 mag) in a larger (8 $''$ radius) aperture to collect all its flux. To estimate the background we randomly placed each aperture in empty regions of the images, taking the sum of the counts. This procedure is repeated 1000 times and we take an outlier clipped average as the best estimator for the background value \citep[see][for more details]{Fossati18}. The flux of the calibration star is converted into a zero point for each image which in turn is used to calibrate the photometry of tick-tock. We propagated the uncertainties on the zero point and on the flux of the inner 3$''$ of the galaxy to constrain the uncertainties of the g-magnitude values. In addition to the points analyzed in \citet[][]{ticktock} we included the few additional photometric points released in the 11$^{\rm th}$ data release. 

Only three observations taken with REM before Modified Julian Day (MJD) $\sim$59650 have been considered science-grade (and therefore reported in figure~\ref{fig:photometry}). The other images have been discarded due to either bad seeing conditions, or too strong a contamination from the Moon light. The second set of REM observations (starting from MJD 59666) have been taken every night, with four observations each night with small shifts for dithering of $<5''$. The points referring to this second set of observations in figure~\ref{fig:photometry} are obtained by merging the four observations of each night. The data of the night MJD 59680 were too affected by the moon light, and have therefore been discarded. The mean and standard deviation of the seeing for the science grade night are 1.21 and 0.24, respectively. The median  sampling data of ZTF is 2.07 days, while for the newly observed data is 1.014 days.
The details of the observing conditions of our REM observations and the values of the derived magnitude and the associated uncertainty are listed in Table \ref{tab:obs}.

\begin{table}
    \caption{Details of the observing conditions and magnitude values of the REM observations, except for the first two epochs which have been observed with NTT. Column 1: average MJD of the co-added exposures for each epoch, column 2: average seeing Full Width at Half Maximum (FWHM) in arcsec of the co-added exposures for each epoch measured by the observatory DIMM probe, column 3: Fractional Lunar Illumination (FLI) at average MJD, column 4: derived $g-$band AB magnitude, column 5: uncertainty on the $g-$band magnitude.}
    \centering
    \begin{tabular}{lcccc}
         \hline
         MJD & Seeing & FLI & Mag & Mag Err  \\
         \hline
            59614.34 (NTT)  & 1.77 & 0.11  &  17.17  &  0.08 \\
            59615.38 (NTT) & 0.81 & 0.19 &  17.18  &  0.09 \\ 
            59619.31  & 1.05 & 0.57 &  17.26  &  0.23 \\  
            59639.38  & 1.09 & 0.03 &  17.54  &  0.34 \\  
            59645.38  & 1.07 & 0.22 &  17.65  &  0.33 \\     
            59666.36  & 1.01 & 0.18 &  17.52  &  0.17 \\     
            59667.26  & 0.67 & 0.11 &  17.38  &  0.08 \\ 
            59668.28  & 1.31 & 0.05 &  17.52  &  0.11 \\     
            59669.38  & 1.01 & 0.01 &  17.71  &  0.12 \\ 
            59670.38  & 0.82 & 0.00 &  17.52  &  0.11 \\    
            59671.26  & 1.43 & 0.01 &  17.38  &  0.11 \\    
            59673.22  & 1.51 & 0.09 &  17.48  &  0.21 \\    
            59674.23  & 1.27 & 0.15 &  17.42  &  0.14 \\    
            59675.24  & 1.17 & 0.23 &  17.50  &  0.11 \\    
            59676.26  & 1.57 & 0.31 &  17.53  &  0.17 \\     
            59677.33  & 1.17 & 0.41 &  17.34  &  0.12 \\    
            59678.33  & 1.17 & 0.51 &  17.52  &  0.10 \\    
            59679.34  & 1.18 & 0.60 &  17.62  &  0.16 \\    
            59681.37  & 1.00 & 0.78 &  17.46  &  0.11 \\ 
         \hline
    \end{tabular}

    \label{tab:obs}
\end{table}

Our photometry campaign demonstrates that the tick-tock nucleus has fainted during the last observations (MJD$\geq 59600$), by $\approx 0.2$ magnitudes, when including all the ZTF data and all the photometric points from our NTT and REM follow-ups (the average dimming when only the second daily set of more accurate REM observations is considered is $\approx 0.26$ mag). Thanks to the small uncertainties of the NTT observations we can observe that the decrease in luminosity was not complete at MJD 59615. No clear modulation is observed in the second set of REM observations, indicating that either no coherent oscillations are present of that such oscillations have decrease significantly in time. Both a decrease of the oscillations and the average dimming of the nucleus are in contrast with the simple binary model discussed in \citep[][]{ticktock}, once we account
for the corrections discussed at the end of next section.

\section{MBHB model description}\label{sec:binary}

In this section, we review the binary model discussed in \cite{ticktock} for the tick-tock MBHB candidate.  

Tick-tock shows a changing-look behaviour, with a significant increase of optical and infrared luminosity (by more than one magnitude in the different bands) over few years and the brightening of broad hydrogen Balmer lines \citep[][]{ticktock}. The early low-luminosity state (low-state hereafter), featuring a faint and blue-shifted H$\alpha$ broad line, has been associated with the secondary MBH being the only active member of the binary, while the brightening of the continuum and of the broad Balmer lines have been associated with the fraction of the material initially surrounding the secondary that accretes onto the primary triggering the continuum emission. 
Indeed, the different MBH mass estimates obtained through the single-epoch method, from spectra taken during the two different accretion states, have been used in support of the binary model by \cite{ticktock}.

The modulation of the continuum with decreasing amplitude and frequency, observed after the brightening of the nucleus (i.e. in the high-state), has been associated with the recurring impacts onto the primary accretion disc or the secondary at its pericentres.
Since the accretion disc is assumed to be geometrically thin/dynamically cold, as long as the secondary is not orbiting on circular velocities in the plane of the disc each impact will be highly supersonic, resulting in strong shocks \citep[see, e.g.][]{Lehto96}.
In order to reproduce the time series of the  peaks in the light curve, an extremely eccentric unequal mass binary with extremely close  pericentres is required \citep[$p\sim 10$ $R_{\rm g}$, where $R_{\rm g}=G M_{\rm BH}/c^2\approx 10^{-4}$ pc for a primary mass $M_1\approx 2 \times 10^8 \msun$, see][]{ticktock}. With such small pericentres the GW emission leads to the shortening of the period between two peaks hinted by the data. The largest apocentre in the \cite{ticktock} model is $\sim 600 - 800 \, R_{\rm g}$, implying an extremely large binary eccentricity at the beginning of the observed modulation.

\citet[][]{ticktock}, built a post-Newtonian (PN) trajectory model to fit the model to their photometric data. Since both the re-analyzed data used in \cite{ticktock} and the numerical analysis have not been released, in order to have a quantitative control on the model we performed independently the same exercise. 

More specifically, the PN model is based on the two-body equations of motion in the Lagrangian formalism derived in \citet{Blanchet2014}.
We consider the PN corrections up to 3.5 PN order beyond Newtonian dynamics, where 1,2,3 PN represent conservative terms, while 2.5 and 3.5 PN orders are dissipative terms that encode the emission of GWs by the inspiralling binary. We adopt the equations describing the relative system, i.e. we evolve forward in time the relative separation $\mathbf{r}$ and velocity $\mathbf{v}$ of the binary according to:
\begin{equation}
    \frac{d^2 \mathbf{r}}{dt^2} = -\frac{G M}{r^2} \biggl( (1+\mathcal{A}) \ \mathbf{n} + \mathcal{B} \mathbf{v} \biggr) + \mathcal{O}\left(\frac{1}{c^8}\right), 
\end{equation}
where $\mathbf{n} = \mathbf{r}/|\mathbf{r}|$, while the coefficients $\mathcal{A}$ and $\mathcal{B}$ can be found in \citet{Blanchet2014}.

We implemented the trajectory model into a Likelihood function used by the Bayesian framework PyMultinest \citep{Buchner2014} in order to fit the luminosity peaks with the pericentres of a binary orbit. A Bayesian nested sampling  algorithm \citep{skilling} has been preferred to a Monte Carlo Markov Chain algorithm since it includes a criterion for convergence that mitigates the risks of a too low number of sampling points. 

Through the sampling, we want to constrain the following five parameters: the masses of the binary components $M_1, M_2$, the orbital eccentricity $e$, the binary pericentre $p$\footnote{We employ the Newtonian relation between eccentricity and  pericentre. Although this does not translate into the true effective pericentre of the binary, it still captures the features of the system involved.} and an offset time $\Delta t$ used to avoid a dependence on the specific origin of the binary evolution time. 
The priors assumed are summarized in table~\ref{tab:stats}.
We stress that, differently from \citet{ticktock}, we stick to the assumption of a bound binary system, i.e. $e<1$, see the discussion in the next section.

\begin{table}
    \centering
    \caption{Priors (column 2) and posterior statistical properties of the model parameters. Columns 3 to 5 report the 25th, the 50th and the 75th percentile, whereas the set of parameters corresponding to the maximum likelihood is shown in column 6.}
    \begin{tabular}{cccccc}
    \hline
    parameter & prior & 25th & 50th & 75th & $\mathcal{L}_{\rm max}$ \\
    \hline\\
    $\log_{10} \dfrac{M_1}{\rm M_\odot}$ & U$(7,9)$ & 7.654 & 7.851 & 8.070 & 7.679 \\
    $\log_{10} \dfrac{M_2}{\rm M_\odot}$ & U$(7,9)$ & 7.842 & 8.018 & 8.183 & 7.524 \\
    $e$ & U$(0,0.9999)$ & 0.955 & 0.969 & 0.978 & 0.987 \\
    $p [R_{\rm g}]$ & U$(13,50)$ & 13.868 & 14.583 & 15.506 & 13.462 \\
    $\Delta t$ [day] & U$(0,3650)$ & 280.789 & 495.592 & 918.363 & 217.092 \\
    \hline
    \end{tabular}
    \raggedright \textbf{Notes:} U$(a,b)$ in column 2 refers to a uniform distribution between $a$ and $b$ for the corresponding parameter.
    \label{tab:stats}
\end{table}

The black points in Figure~\ref{fig:photometry} represent the original ZTF data. 
We fitted each data peak with a Gaussian profile, to estimate the peak time and its uncertainty, i.e. the $\sigma$ of the Gaussian. We then used the peak times and uncertainties as input parameters for our Bayesian analysis.

\begin{figure*}
    \centering
    \includegraphics[scale=0.5]{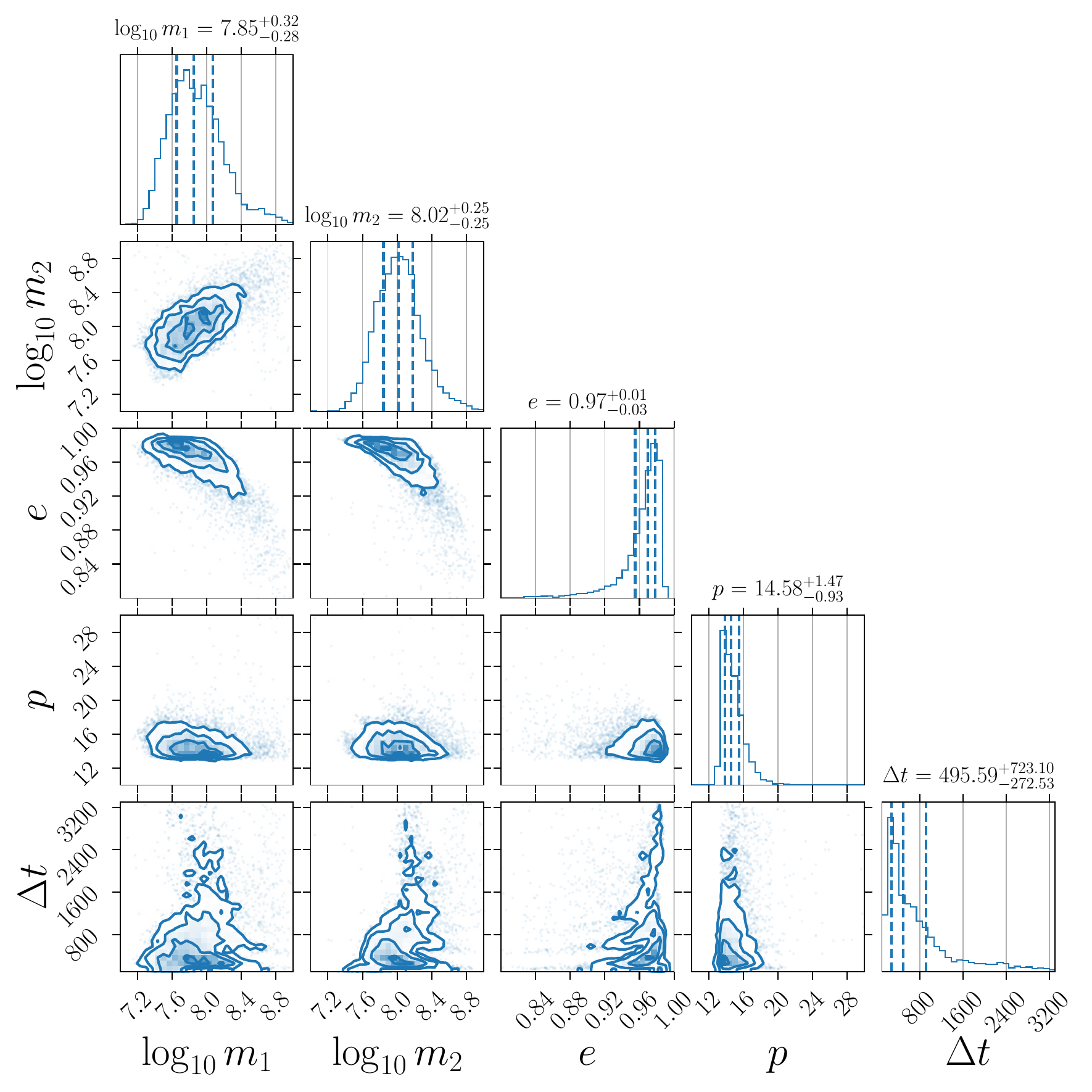}
    \caption{Corner plot of the model parameters with posterior distribution at the top of each column. Marked values denote the median inferred for each parameter, while vertical dashed lines represent the 25th, 50th and 75th quantile.}
    \label{fig:corner}
\end{figure*}

We used only the peaks originally observed in the optical bands, since the inclusion of the UV and X-ray data results in significantly poorer fits, as shown in Figure~3 of \citet[][]{ticktock}. \footnote{As the authors stressed, the worsening of the fit when including the higher frequency photometric points could be due to the natural variability of the inner regions of the accretion disc and of the corona.} The blue curve in Figure~\ref{fig:photometry} shows the orbit with the maximum likelihood found by our nested sampling algorithm, while the cyan lines show the next 50 orbits with highest likelihood. The full posterior distributions for the sampled parameters are shown in figure~\ref{fig:corner}, while the statistical indicators of the marginalized posteriors of each parameter are summarized in table~\ref{tab:stats}.  

\citet{ticktock} derived two predictions from the MBHB model: (1) at the binary separations corresponding to the observed data they predict that the the amplitude of the flux fluctuations should decrease in time, due to the increase of the impact velocity of the secondary onto the primary accretion disc \citep{ticktock}; (2) at smaller separations, corresponding to secondary-disc impacts  closer and closer in time, the observed flux is predicted to increases as the energy released by the impacts fails to thermalize between two consequent impact \citep[][]{ticktock}. The first prediction is, however, flawed by an error in on of its assumptions.

Most of the binary energy is lost in GW close to the pericentres, resulting is close to constant pericentres and decreasing apocentres (as shown in Figure~\ref{fig:photometry}), hence promoting the circularisation of the binary. The impacts between the eccentric secondary and the primary accretion disc (immediately before and after the pericentres) would then occur at roughly constant radii and with  decreasing impact velocities \footnote{Note that, as previously commented, the velocities would remain highly supersonic during the whole evolution.}, \citep[contrary to the original assumption in ][]{ticktock}. Once we account for these corrections in the \cite{ticktock} model, the amplitude of the modulation is, if anything, predicted to increase with time. 

\section{Constraints on the MBHB hypothesis}\label{sec:tests}

Our photometry campaign demonstrates that, during the last observations (MJD$\geq 59600$), the tick-tock nucleus has fainted by $\approx 0.2$ magnitudes (when including all the ZTF data and all the photometric points from our NTT and REM follow-ups), whereas the average dimming is $\approx 0.26$ mag when only the second daily set of more accurate REM observations is considered. Thanks to the small uncertainties of the NTT observations, we can see that the decrease in luminosity was not complete at MJD 59615. No clear modulation is observed in the second set of REM observations, indicating that either no coherent oscillations are present or that such oscillations have significantly decreased their amplitude in time. 
The decrease of the oscillations (already visible in the light curve presented in \citealt{ticktock}) and the average dimming of the nucleus (most clearly visible with the new photometric points) are both in contrast with the simple binary model proposed, once the correct evolution with time of the binary relative velocity at pericentre is taken into account, as discussed in the previous section. The possibility of a binary interpretation of the dimming in the light curve of the binary candidate OJ287 has been discussed in \cite{Lehto96}. In their model a dimming could be associated to the cyclic eclipse caused by the passage of the secondary along the line-of-sight. The authors however comment that such an explanation is plausible only if the main responsible of the optical emission is the jet emitted by the primary, since the optically emitting region of the primary accretion disc would be to large to be significantly affected. The broad-band spectral energy distribution of tick-tock does not show a significant jet contribution \citep{ticktock}, although a sub-pc radio emission for the object has been detected in \citet{2022A&A...663A.139A} and associated to either a compact radio jet or the AGN corona. The negligible contribution of the jet to the optical bands strongly disfavours this interpretation for the specific source under scrutiny.
Although we are aware that the proposed model is very idealised and possibly oversimplified \citep[see, e.g.][for a numerical study including a richer and more complex dynamics]{Bode2012}, the photometric points disfavour the binary scenario. The most plausible interpretation is that the observed dimming is part of the AGN variability observed even before the appearance of the quasi-periodic modulation, and not associated with a MBHB.

In addition to the arguments based on the observed light curve, we propose some other cautionary comments about the binary interpretation of the tick-tock observational features:

\begin{itemize}

\item[$\bullet$] The moving secondary is a tempting explanation for the H$\alpha$ in the low-state SDSS spectrum. Indeed we repeated the analysis with the publicly available QSFIT code \citep[][]{qsfit} and found consistent results. We stress however that ($i$) the line equivalent width is quite low, and ($ii$) the blue side of the fitted H$\alpha$ could be affected by blended iron transitions and Bowen fluorescence lines, as observed in other type-I AGN \citep[][]{2004-veron-spectra-izw1, benny2}. These two points suggest that some additional caution is needed when evaluating the relevance of the single epoch estimate of the secondary mass $M_2$, that is anyway consistent with the estimated mass of the primary from the later spectra presented in \citep[][]{ticktock} once the scatter in the single epoch relation used \citep[][]{singleepoch} is taken into account. We further stress that similar Balmer profiles have been observed in other proposed changing-look AGN without periodic  modulations, for which no binary model has been proposed \citep[see, e.g. the spectra taken in March and April 2018 of 1ES 1927+654][]{benny}.

\item[$\bullet$]  The extremely high initial eccentricity is hard to reconcile with the proposed three-body scatterings with single stars as stellar (as well as gas) driven hardening acts on a typical timescale of, at best, several millions of years \citep{2010ApJ...719..851S}. When this becomes comparable to the GW coalescence timescale, GWs take over, leading to significant circularisation. While stellar hardening typically leads to eccentricity growth, even considering an extreme case of a MBHB with $e=0.99$ \citep[the highest eccentricities observed in dynamical studies of binary formation, e.g.,][]{2012ApJ...749..147K} 1 Myr from coalescence, GW emission would drastically circularize its orbit by the final years before merger, with $e << 0.1$ at coalescence \citep{2022MNRAS.511.4753G}. 
The only viable explanation discussed in literature requires the presence of a close third MBH (a perturber that is much more massive than single stars). Secular interaction in a hierarchical configuration \citep[as proposed in ][]{ticktock} can possibly explain the bound initial conditions with parameters from the optical variability fit, but would not lead to the very high eccentricity required in the unbound cases preferred by the fits including X and UV photometric points, as generally precession would prevent the occurrence of Lidov-Kozai oscillations\footnote{But see \citet{2013ApJ...773..187N} for conditions where relativistic precession could not suppress eccentricity excitation.} \citep[see e.g.][]{2000ApJ...535..385F,2002ApJ...578..775B,2016ARA&A..54..441N,2016MNRAS.461.4419B,2018PASA...35...17B}. 

\item[$\bullet$] Finally, taking into account the ZTF sensitivity and sky coverage and considering that, in the binary interpretation, tick-tock is expected to coalesce within the next couple of years, \citet{ticktock} estimated a merger rate of $\approx 0.03-0.05\,$yr$^{-1}$Gpc$^{-3}$ for this type of systems. This is in severe disagreement with estimates from models of MBHB hierarchical growth, which predict that the average $z<1$ merger rate for MBHBs with masses of $10^8\msun$ and higher lies somewhere in the 10$^{-5}$--10$^{-4}$yr$^{-1}$Gpc$^{-3}$ range.
This low rate derives from the fact that the MBH mass function is a steeply decreasing function of mass, therefore the merger rate is strongly dominated by low mass systems and when a mass cut at $10^8 \rm M_\odot$ is applied, the rate decreases significantly. In comparison, rates up to 10$^{-2}$yr$^{-1}$Gpc$^{-3}$ are found considering the MBHB mass function down to $10^7 \rm M_\odot$ \citep[][]{2008MNRAS.390..192S}.
Finally, one could of course argue that the sensitivity limit of ZTF would allow to see local AGNs down to a much lower MBH mass, say $10^7 \rm M_\odot$, in the local universe, which would make the rate estimate consistent with expectation. In this case, however, one would have to face the very small odds of seeing only one $10^8 \rm M_\odot$ binary in a population that should be statistically heavily dominated by much lighter systems.
\end{itemize}

\section{Caveats of the current study}\label{sec:caveats}

The main caveat of the analysis presented concerns the time coverage of our photometric follow-ups. Unfortunately, the best fitting orbit we derive from our Bayesian analysis would have a pericentre occurring during the gap in our observations, somewhat limiting the constraining power of the new data. We stress however that, due to the limited number of peaks fitted and to the large number of parameters in the binary orbit, the exact values of pericentres far in time from the original ZTF data cannot be properly constrained. Indeed, a few realizations in our sampling have pericentres occurring immediately before our first and brightest observations. Further follow-ups and other already proposed observational tests to probe the nature of such system could potentially result in a more secure interpretation of its nature \citep{2022A&A...665L...3D,2022MNRAS.509..212D}.

In the discussion of the dynamical processes leading to the formation of an extremely eccentric MBHB we did not considered the possible effect of a third MBH. Indeed, for bound binary orbits as well as for initially unbound ones, the only viable process able to dramatically increase the eccentricity at the level proposed by \citet{ticktock} requires a sudden quasi-radial scatter of the secondary onto the primary prompted by the chaotic interaction with a close third MBH. Such a prompt deviation would be unaffected by the above-mentioned relativistic precession quenching. Although this kind of process may be a physically possible channel, the likelihood of this type of event has not been properly quantified in \citet[][]{ticktock}, and tend to occur with a lower frequency with respect to the other MBHB evolutionary channels \citep{2018MNRAS.477.3910B}.

There are other limitations associated with the simple binary model assumed. In particular, as in \citealt{ticktock} we consider a geometrically thin accretion disc orbiting the primary MBH. Such a thin configuration could be an over-simplification, due to repeated impacts between the disc and the secondary and the recent formation of the newly acquired accretion disc around the primary (see section~\ref{sec:binary}). In addition, the secondary could exert an additional radiative (and possibly mechanical) feedback onto the primary disc. We consider however such last possibility unlikely, since in the model the luminosity of the secondary is assumed to be subdominant \citep[][]{ticktock}.

\section{Conclusions}\label{sec:conclusions}

We presented a discussion of the main observational properties of tick-tock and of the binary model proposed in \citet[][]{ticktock}. We commented about some weaknesses of this model, regarding  the extreme binary parameters needed (that indeed would require fine tuning in the dynamical processes responsible for the binary formation), the limited significance of the active MBH masses estimated from single-epoch spectroscopy at different epochs, and the incorrect prediction of decreasing amplitude of the magnitude fluctuations. When using the correct evolution of the binary pericentre the model predicts an increase in the fluctuation magnitude, while a decreasing trend is observed in the few fluctuations present in the ZTF data. 

We presented the results of a dedicated optical follow-up campaign, that indicate that the AGN luminosity decreased by $\approx 0.2$ magnitude (or more, depending on the sample over which the average is performed), down to values comparable with the luminosities observed until 2018 \citep[see figure~1 in the methods appendix in][]{ticktock}. Due to the limited time-coverage of our campaign and on the uncertainties in the magnitude measurements it is difficult to determine whether the chirping oscillatory trend observed in the ZTF data continued. There is only an indication of an initial decrease in luminosity on a $\sim 20$ day timescale, after which the g-band magnitude remains approximately constant on similar timescales.

Although our analysis cannot fully disprove the MBHB scenario (see the discussion in section~\ref{sec:caveats}), we believe that the extreme model proposed in \citet[][]{ticktock} requires the strongest observational evidence, and that such support is not available for the tick-tock system.

A plausible alternative explanation is that the observed modulations are only due to the intrinsic red-noise of the AGN luminosity of a single MBH \citep[][]{vaughan}. It should also be noted that comparing the
binary model with a red noise model is not straightforward though. Moreover, the comparison would rely on a robust model for the red noise description in AGNs. Despite damped random walk models \citep{2009ApJ...698..895K} or Carma \citep{2014ApJ...788...33K} models have been proposed in the literature, it is unclear which model would best describe the intrinsic random variation of AGNs, therefore, any Bayesian comparison of the binary vs red noise model would in any case depend on the assumed red noise model. Although a detailed Bayesian comparison would be interesting, the main goal of this paper is to demonstrate that the binary model has a
number of potential issues, and the demonstration that this model is more probable than red noise is beyond the scope of this study.

Here we speculate about a third scenario, in which an initially centrally-carved accretion disc misaligned with respect to the MBH spin refills episodically its central regions \citep[as recently observed in MHD simulations; see, e.g.,][]{2019MNRAS.487..550L,2021MNRAS.507..983L}. The transient inner disc would undergo Lense-Thirring precession \citep[e.g.][]{LenThir} around the MBH spin axis with a frequency significantly lower than that of the outer disc \citep{2021MNRAS.507..983L}, resulting in a varying relative orientation of the inner disc with respect to the line of sight and, therefore, a varying observed flux. 
Indeed this model and the red noise interpretation mentioned above may be not exclusive, as a transient precessing disc with evolving precession frequency could be one of the processes contributing to the red-noise observed in AGN.

The correct modeling of a MBH/misaligned-disc precessing system  is not straightforward. While the precession is indeed forced by the frame-dragging exerted by the MBH metric, the precession frequency of the inner disc can be significantly modified by the magnetic torques present in MHD simulations \citep{2021MNRAS.507..983L}. In addition, the analytical form of the evolution of the outer radius of the inner disc (indeed observed in MHD simulations and needed to model the decreasing in amplitude-chirping light curve observed) is not available to date. We postpone the development of a complete model for the single MBH-precessing misaligned disc scenario to a future investigation.

\section*{Acknowledgements}
The Authors acknowledge the valuable help of the staff of REM, an INAF facility managed by the REM team (www.rem.inaf.it). A.F., A.S. and E.B. acknowledge financial support provided under the European Union’s H2020 ERC Consolidator Grant ``Binary Massive Black Hole Astrophysics'' (B Massive, Grant Agreement: 818691). AL acknowledges funding from MIUR under the grant PRIN 2017-MB8AEZ.

\section*{Data Availability}

The data underlying this article will be shared on reasonable request to the corresponding author.



\bibliographystyle{mnras}
\bibliography{biblio} 


\bsp	
\label{lastpage}

\end{document}